\documentclass[epjCONF]{svjour}
\usepackage{graphics}
\usepackage[varg]{txfonts} 
\usepackage[latin1]{inputenc}
\usepackage{xspace}

\session-title{Conference Title, to be filled}

\begin{document}

\def\Kepler{\textit{Kepler}\xspace}
\def\Hermes{\textsc{Hermes}\xspace}
\def\gpsc{$\gamma$\,Psc\xspace}
\def\ttau{$\theta^1$\,Tau\xspace}

\title{Oscillating red-giant stars in eccentric binary systems
}
\author{P.\,G. Beck\inst{1,2}\fnmsep\thanks{\email{paul.beck@cea.fr}} 
\and K. Hambleton\inst{3,2} 
\and J. Vos\inst{2}
\and T. Kallinger\inst{4}
\and R. A. Garcia\inst{1}
\and S. Mathur\inst{1}
\and K. Houmani\inst{1}}
\institute{Service d'Astrophysique, IRFU/DSM/CEA Saclay. 
\and Instituut voor Sterrenkunde, KU Leuven, Belgium.
\and University of Central Lancashire, Preston UK. 
\and Institut f\"ur Astronomie der Universit\"at Wien, Austria. 
}
\abstract{
The unparalleled photometric data obtained by NASA's Kepler Space Telescope has led to improved understanding of red-giant stars and binary stars. We discuss the characterization of known eccentric system, containing a solar-like oscillating red-giant primary component. We also report several new binary systems that are candidates for hosting an oscillating companion. A powerful approach to study binary stars is to combine asteroseimic techniques with light curve fitting. Seismology allows us to deduce the properties of red giants. In addition, by modeling the ellipsoidal modulations we can constrain the parameters of the binary system. An valuable independent source are ground-bases, high-resolution spectrographs. 
} 
\maketitle
\section{\vspace{-2mm}Red-giant heartbeat stars}
The observation of a vast number of objects observed with the NASA \Kepler space telescope \cite{borucki} has led to numerous fascinating findings. Among others a new type of ellipsoidal modulated binary stars were found. In these eccentric binary systems, colloquially referred to as $heartbeat$ stars, a modulation of the luminosity at the periastron is induced by tidal interaction between the stars \cite{welsh}. For red-giant stars, the exquisite photometric data quality from the space mission allowed us to better understand the oscillation spectrum and even investigate the core rotation of these elderly stars \cite{beck2012}.

In this poster we discussed a sample of 18 eccentric binary systems with an oscillating red-giant star as primary component \cite{beck2014a}. Within this sample, the red-giant star with the richest power spectrum is  KIC\,5006817. It was subject of a combined analysis approach: Seismology reveals a red-giant primary of $\sim$6R and $\sim$1.5M and, by modeling the effects of rotation \cite{beck2012}, it can be shown that in this star the core rotates 13-times faster than the surface, which takes $\sim$165 days for a revolution. This is twice the orbital period and excludes pseudosynchronisation. Spectroscopy and light curve modeling proved that the system is eccentric (e=0.72) and that the companion is likely to be a G dwarf with $\sim$0.3M$_\odot$. Modeling the binary evolution shows that the system will undergo a common envelope phase once the expanding radius reaches $\sim$100R. Moreover, Doppler beaming must have a significant contribution, but it is hidden in the instrumental noise. The study of KIC5006817 shows the high potential of the combined analyses, using seismology and light curve modeling. It, in fact, leads to a comprehensive picture of KIC5006817, including strong constraints on the invisible secondary companion. 

\begin{figure}[t!]
\centering
\resizebox{0.80\columnwidth}{!}{  \includegraphics{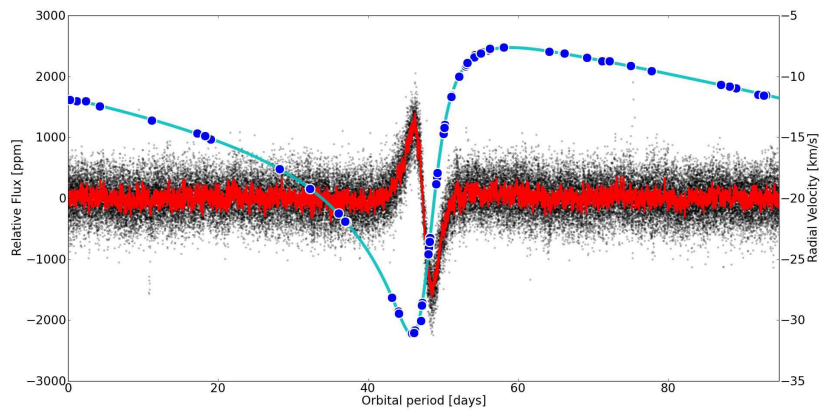} }
\caption{\label{fig:kic50068}
Combined phase diagram for the eclipsing system KIC\,5006817 \cite{beck2013PhD}  of space photometry by the \Kepler satellite and radial velocities from ground-based spectroscopic observations with the \Hermes spectrograph. The \Kepler photometry is presented on the left vertical axis. Black data points are the original measurements, while the red line shows the average lightcurve, which was determined by rebinning the phase diagrams into 30 minutes bins. The deviation from the average light curve is caused by solar-like oscillations. The radial velocities are shown as solid, blue dots (right vertical axis). The solid blue line represents the orbital model of a Keplerian orbit. }
\end{figure}

\section{\vspace{-2mm}Radial velocities: Orbital parameters and solar-like oscillations}
A major ingredient to solve these systems is the determination of the orbital parameters from fitting a orbital model to the radial velocities derived from ground based spectroscopy. Radial velocities allow to set an independent constraint on the orbital eccentricity, putting strong constraints on the light curve model and narrowing down the space of free parameters. We therefore monitored the 18 binary systems \cite{beck2014a} with about two to three dozens observations each with the \textsc{Hermes} spectrograph \cite{raskin} within a year. 

The derived orbital eccentricities range from 0.2 to 0.76 with orbital periods from 20 to 440 days. While their masses and radii do not deviate from the sample of red giants observed with \Kepler, it seems that all stars in the sample are in the less advanced hydrogen-burning phase. This supports the argument that these systems either eject their hull after undergoing a common envelope phase or circularize as the red giant ascends the red-giant branch.

We note, that radial velocities of red-giant stars can reveal even more. If an instrument can be calibrated to the level of meters-per-second, such as \textsc{Hermes}, solar-like oscillations are also detectable from ground-based high-resolution spectroscopy. Recently, $\theta^1$\,Tau, a bright red-giant star in a long periodic binary system was the target of an extensive spectroscopic observing campaign \cite{beck2014b}, allowing to confront seismic mass and radius with independent literature values. A good agreement was found.

\section{\vspace{-2mm} Known and new heartbeat stars}
While working on compiling \Kepler light curves from individual quarters \cite{garcia}, we found new \textit{heartbeat} systems. The systems KIC\,10322133, KIC\,10188415, and KIC\,6042423 show signal in the frequency range above 100\,$\mu$Hz, which could originate from oscillations. Non oscillating, but interesting due to its long period of more than 400\,days, is KIC\,10460156. Finally, we report KIC\,10074700 as a new eclipsing binary with solar-like oscillations. The seismic and orbital characteristics of these newly found systems will be explored in a subsequent work. 

\vspace{-2mm}

\end{document}